\documentclass[prb,twocolumn,amssymb,amsmath,nobibnotes]{revtex4}

\usepackage{graphicx,color}
\usepackage{subfigure}
\usepackage{dcolumn}
\usepackage{bm}
\usepackage{threeparttable}
\usepackage{stmaryrd}
\begin{document}

\newenvironment{figurehere}
  {\def\@captype{figure}}
  {}

\title{Adsorbate and defect effects on electronic and transport properties of gold nanotubes}
\author{Yongqing Cai$^1$}
\author{Miao Zhou$^1$}
\author{Minggang Zeng$^1$}
\author{Chun Zhang$^{1,2}$}
\email{phyzc@nus.edu.sg}
\author{Yuan Ping Feng$^1$}
\email{phyfyp@nus.edu.sg}
\affiliation{
	$^1$Department of Physics, National University of Singapore,
		2 Science Drive 3, Singapore 117542\\
	$^2$Department of Chemistry, National University of Singapore}
\date{\today}

\begin{abstract}
 First principles calculations have been performed to study the effects of adsorbates(CO molecule and O atom) and defects on electronic structures and transport properties of Au nanotubes (Au(5,3) and Au(5,5)). For CO adsorption, various adsorption sites of CO on the Au tubes were considered. Vibrational frequency of the CO molecule was found to be very different for two nearly degenerate stable adsorption configurations of Au(5,3), implying the possibility of distinguishing these two configurations via measuring the vibrational frequency of CO in experiment. After CO adsorption, the conductance of Au(5,3) decreases by 0.9 $G{_0}$, and the conductance of Au(5,5) decreases by approximately 0.5 $G{_0}$. For O adsorbed Au tubes, O atoms strongly interact with Au tubes, leading to around 2 $G{_0}$ of drop of conductance for both Au tubes. These results may have implications for Au-tube based chemical sensing. When a monovacancy defect is present, we found that for both tubes, the conductance decreases by around 1 $G{_0}$. Another type of defect arising from the adhesion of one Au atom is also considered. For this case, for Au(5,3) tube, it is found that the defect decreases the conductance by nearly 1 $G{_0}$, whereas the decrease of the conductance is only 0.3 $G{_0}$ for Au(5,5) tube due to the adsorption of the extra Au atom.
\end{abstract}

\maketitle

\section{INTRODUCTION}

   Gold quasi one-dimensional (1-d) nano structures, such as nanowires and nanotubes, have attracted lots of attention due to their intriguing physical and chemical properties that are very different from bulk gold.\cite{Y. Kondo,STM,Break Junction,Y. Oshima_prl} The recent rapid progress of experimental techniques enables us to fabricate these gold quasi 1-d structures and measure their novel properties. Long gold monoatom chains have been produced in experiments by depositing single Au atoms onto a metallic NiAl(110) surface using a scanning tunneling microscopy(STM) tip,\cite{N. Nilius_science,N. Nilius_prl} and the unit conductance of such gold chain was reported by another experiment.\cite{A. I. Yanson} Gold nanowires suspended between two bulk electrodes have been fabricated and intensively studied by either STM techniques \cite{STM} or the break junction method.\cite{Break Junction} More recently, helical single-wall gold nanotubes were successfully synthesized in experiment.\cite{Y. Oshima_prl} Theoretical studies have shown that among all freestanding gold tubes, Au (5,5) is the most stable one, and when suspended between two gold electrodes, the Au tube (5,3) is the most energetically favored.\cite{R. T. Senger}

   Owning to their helical structures, gold nanotubes have unique electronic and catalytic properties. It was predicted by theoretical study that the chiral current flowing through gold tubes may induce strong magnetic field; \cite{Spiral_Current,D. Z. Manrique} Moreover, both experimental and theoretical investigations have demonstrated excellent catalytic activity of gold nanotubes.\cite{M. A. Sanchez-Castillo,W. An} These unique properties of gold nanotubes suggest promise of gold tubes for future applications of nanoelectronics and nanocatalysis. In this paper, we shall discuss effects of two important factors that may have great implications for real applications, chemical modifications and defects, on electronic and transport properties of gold tubes.

   The chemical modification, in particular, the adsorption or doping of small molecules, has been regarded as a very effective way to tune and control the electronic structure and conductance of quasi 1-d nano  structures.\cite{Zhang_Nanowire} For instance, for In nanowires, the adatoms of O and In can individually suppress more than one third of the conductance of the wire.\cite{S. Wippermann} and  the adsorption of a single CO atom on Au nanochain turns the system from metal to semiconductor.\cite{N. Nilius_prl,A. Calzolari}For Au nanotubes, the interaction of molecules and atoms with the tubes still needs to be clarified. Defects are important because first, in the procedure of fabrication, in principle, defects are inevitable, and second, defects have been found to have significant effects on physical and chemical properties of nano systems, which in many cases are essential for applications, for example, greatly defect-enhanced catalytic activity of gold nano clusters on graphene. \cite{ZMpaper} In the present study, the adsorption of CO molecule and O atom, and two types of defects, Au monovacancy and the defect arising from the adsorption of one extra Au atom on the tube, are considered.

\section{COMPUTATIONAL DETAILS}
   For structural optimization and electronic structure calculations, the first principles method based on density functional theory (DFT) was employed via the computational package VASP.\cite{VASP_jcm,VASP_prb} During calculations, a plane wave basis set with the cut-off energy 396 eV was used. The structures were relaxed until the force less than 0.02 eV/${\AA}$. The transport properties were calculated by ATK code within the nonequilibrium Green's function (NEGF) formalism.\cite{ATK_1,ATK_2,ATK_3} Double-$\zeta$ polarized basis and a cutoff energy of 150 Ry for the grid integration were adopted in transport calculations. In all calculations, the Perdew-Burke-Ernzerhof(PBE) format of GGA approximation was included.\cite{PBE_GGA} Vibrational analysis was done using DMol$_{3}$,\cite{DMol_1,DMol_2} and the effective core potential (ECP) and a double numerical basis set including a d-polarization function (DND) were adopted.

\section{RESULTS AND DISCUSSION}

  \subsection{Adsorption configurations of CO, O and Au on Au(5,3)/(5,5) tubes}

   Fig. 1 shows atomic structures of Au(5,3)/(5,5) tubes and possible adsorption sites (top, center and two inequivalent bridge sites, B1, B2) of CO molecule or O, Au atoms. In Table I, we list the adsorption energy, bonding length defined as the shortest distance between the adsorbate (CO, O and Au) and the nearest Au atom, and the change of net charge of the adsorbates and the bonded Au atoms, for most stable adsorption configurations of different cases. In the case of Au(5,3), for each adsorbate, we found that the difference between the binding energy of the lowest-energy configuration(top site adsorption) and that of the second lowest one(B1 site adsorption) is around 0.03 eV. These two adsorption configurations may co-exist in experiments. We therefore show both of them in the table. Other adsorption configurations (not shown in the table) have binding energies at least 0.4 eV lower than the most stable one.

\begin{figure}
  \includegraphics[width=8.0cm]{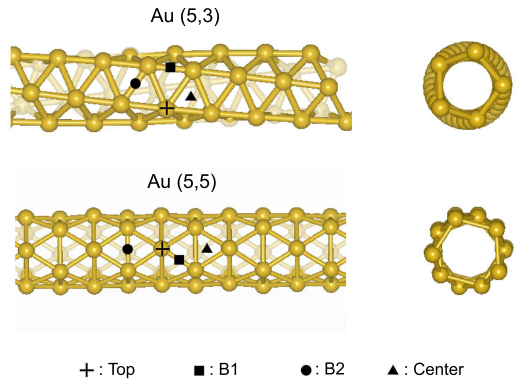}\\
  \caption{(Color online) Geometric structures of (5,3) and (5,5) Au nanotubes and a schematic description of possible adsorbing sites for adsorption: Top, B1 (Bridge site 1), B2 (Bridge site 2), Center.}\label{Fig.1}
\end{figure}

\begin{table}[htbp]
 \centering\small
 \begin{threeparttable}
 \caption{\label{tab:results}Energetics and structures of adsorbates(CO, O, Au) on Au(5,3) and Au(5,5) nanotubes. The adsorption energy $\emph{$\Delta$}E$ is calculated by subtracting the energy of the total system from the sum of energies of the nanotubes and the adsorbates; $\emph{d}$ is the shortest length of the bonds formed by Au and adsorbates; \emph{$\delta$Q$_{x}$}(\emph{x}=CO, O, Au) and \emph{$\delta$Q$_{Au}$} are the net partial charge transfers of the adsorbates and the Au atom with the shortest distance from the adsorbates.}
  \begin{tabular*}{0.5\textwidth}{@{\extracolsep{\fill}}lcccc}
  \hline
  \hline
  Adsorbate  &    $\emph{$\Delta$}E$(eV)(site)  &   $\emph{d}$({\AA})    &     \emph{$\delta$Q$_{x}$} & \emph{$\delta$Q$_{Au}$}  \\   
  \hline
      &                &    Au(5,3)  &           &          \\
   CO &  0.57(top)     &    1.99     &  0.389    &  -0.090  \\
      &  0.54(B1)      &    2.12     &  0.399    &  -0.045  \\
   O  &  4.66(center)  &    2.12     &  -0.522   &  0.150   \\
      &  4.65(B1)      &    2.03     &  -0.528   &  0.170   \\
   Au &  2.28(center)  &    2.73     &    -0.060 &  0.028   \\
      &  2.27(B2)      &    2.71     &   -0.048  &  0.009   \\
  \hline
      &                &    Au(5,5)  &           &          \\
   CO &  0.77(B1)      &    2.10     &   0.441   &  -0.050  \\
   O  &  5.13(center)  &    2.05     &  -0.507   &  0.139   \\
   Au &  2.69(center)  &    2.74     &   -0.006  &   0.045  \\
  \hline
  \hline
 \end{tabular*}
 \end{threeparttable}
\end{table}

   The side views of optimized geometries of CO or O adsorbed Au(5,3)/(5,5) tubes are shown in Fig. 2. For the case of CO adsorbed Au(5,3), two nearly degenerate stable adsorption configurations are found as above mentioned: The top site adsorption (Fig.2a) where the C atom directly binds to the Au atom, and the B1 adsorption (Fig.2b) where the CO is on the bridge site of Au chains along the direction parallel to the tube axis. For the top site adsorption, the C-Au bonding length is 1.99${\AA}$ comparable to that of CO adsorption on Au chain(C-Au bond length 1.96 ${\AA}$) on NiAl surface.\cite{A. Calzolari} As shown in Fig. 2a, after the CO adsorbed on top site, the Au(5,3) tube undergoes a significant distortion, clearly suggesting the strong interaction between the CO molecule and Au tube. For the B1 site adsorption (Fig.2b), the distortion of the tube is much weaker although the adsorption energy in this case is almost the same as the top site adsorption. For the case of Au(5,5) tube, the lowest energy state occurs when the CO molecule binds to the B1 site (Fig. 2c). The binding energy, 0.77 eV, is higher than the case of Au(5,3) by about 0.2 eV. In all cases, electron transfers from the CO molecule to the adjacent Au atom, leading to slightly positively charged  CO molecule and negatively charged Au atom (Table I).

\begin{figure}
  \includegraphics[width=7.0cm]{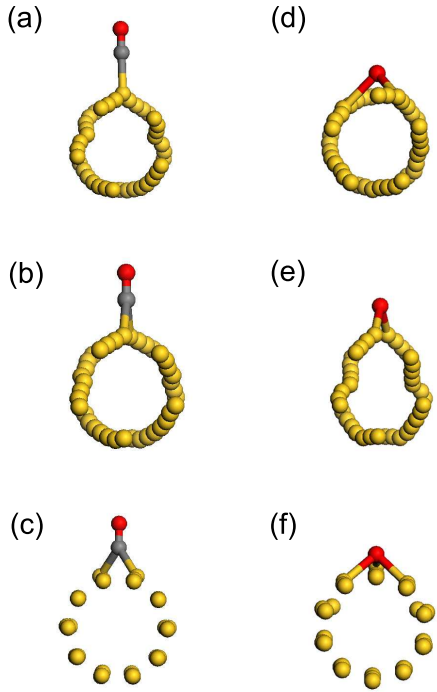}\\
  \caption{(Color online) Adsorption geometries for CO adsorption on Au(5,3) at top (a), B1 (b), and Au(5,5) at B1 (c) sites, oxygen adsorption on Au(5,3) at center (d), B1 (e), and Au(5,5) at center (f) sites, respectively. The C, O, and Au atoms are depicted in grey, red, and yellow, respectively. }\label{Fig.2}
\end{figure}

   Compared to CO, the binding of O atom on Au tubes is much stronger (Table I). For the case of Au(5,3), similarly, there are two nearly degenerate adsorption configurations, the center (Fig. 2d) and the B1 site (Fig. 2e) adsorption, with the adsorption energy 4.66 and 4.65 eV respectively. For Au(5,5), the center site adsorption (Fig. 2f) is the most stable configuration with the adsorption energy of 5.13 eV. When binded on B1 site of Au(5,3), a significant distortion of the Au tube occurs. Moreover, a relatively larger electronegativity of O atom enables a net partial charge transfer from Au to the adsorbate and makes gold atoms positively charged (Table I).

   The adsorption of one extra Au atom on Au tubes is also considered. For both Au tubes, the adsorption energy of the lowest energy configuration is larger than 2.2 eV (Table I), suggesting that in the procedure of fabrication, it is very possible to have Au adatoms on Au tubes. Later, we will show how the Au adatom affects electronic and transport properties of Au tubes.
  \subsection{Electronic structures of CO and O adsorbed Au(5,3)/(5,5) tubes}

  Here, we describe the electronic structures of CO/O absorbed Au(5,3) and Au(5,5) tubes. First, to understand the interaction between CO molecule and Au tubes, we plot in Fig. 3 local density of states (LDOS) projected on the CO molecule and the binded Au atom. Similar to the case of CO binding to Au nano clusters,\cite{Landman_science} the highest occupied molecular orbital(HOMO) and the lowest unoccupied molecular orbital(LUMO) of CO have the biggest contribution to the chemisorption since the energy levels of these two states change significantly upon adsorption due to the hybridization with Au states. For top site adsorption on Au(5,3) (Fig. 3a), the CO HOMO orbital $5\sigma$ is pushed below the $1\pi$ orbital whose energy level doesn't change much, and the LUMO orbital $2\pi^{*}$ is pushed below Fermi energy and overlap with the whole continuous $d$ band of Au, leading to the partial population of this orbital (the backdonation process \cite{G. Blyholder,M. Kiguchi}). The amount of electrons backdonated to the $2\pi^{*}$ orbital can be roughly estimated by integrating the LDOS of CO within the energy range of the Au $d$ band. In this case, about 0.9 electron is transferred from Au $d$ bands to the CO anti-binding orbital $2\pi^{*}$. In the figure, we also plot the wave functions of several molecular orbitals of CO after adsorption, from which we can see that indeed the $5\sigma$ orbital strongly hybridize with Au states, and $4\sigma$ as well as two degenerate $1\pi$ orbital ($1\pi-1$ and $1\pi-2$) only weakly interact with the Au tube. When CO adsorbs on B1 site of Au(5,3) (Fig. 3b), differently from top site adsorption, $1\pi$ orbital also contributes to the chemisorption by significantly hybridizing with Au states, and as a result, the degeneracy between $1\pi-1$ and $1\pi-2$ is lifted. Amount of electrons backdonated to $2\pi^{*}$ orbital in this case (1.57 electrons) is estimated to be much larger than that of the top site adsorption.

\begin{figure}
  \includegraphics[width=8.0cm]{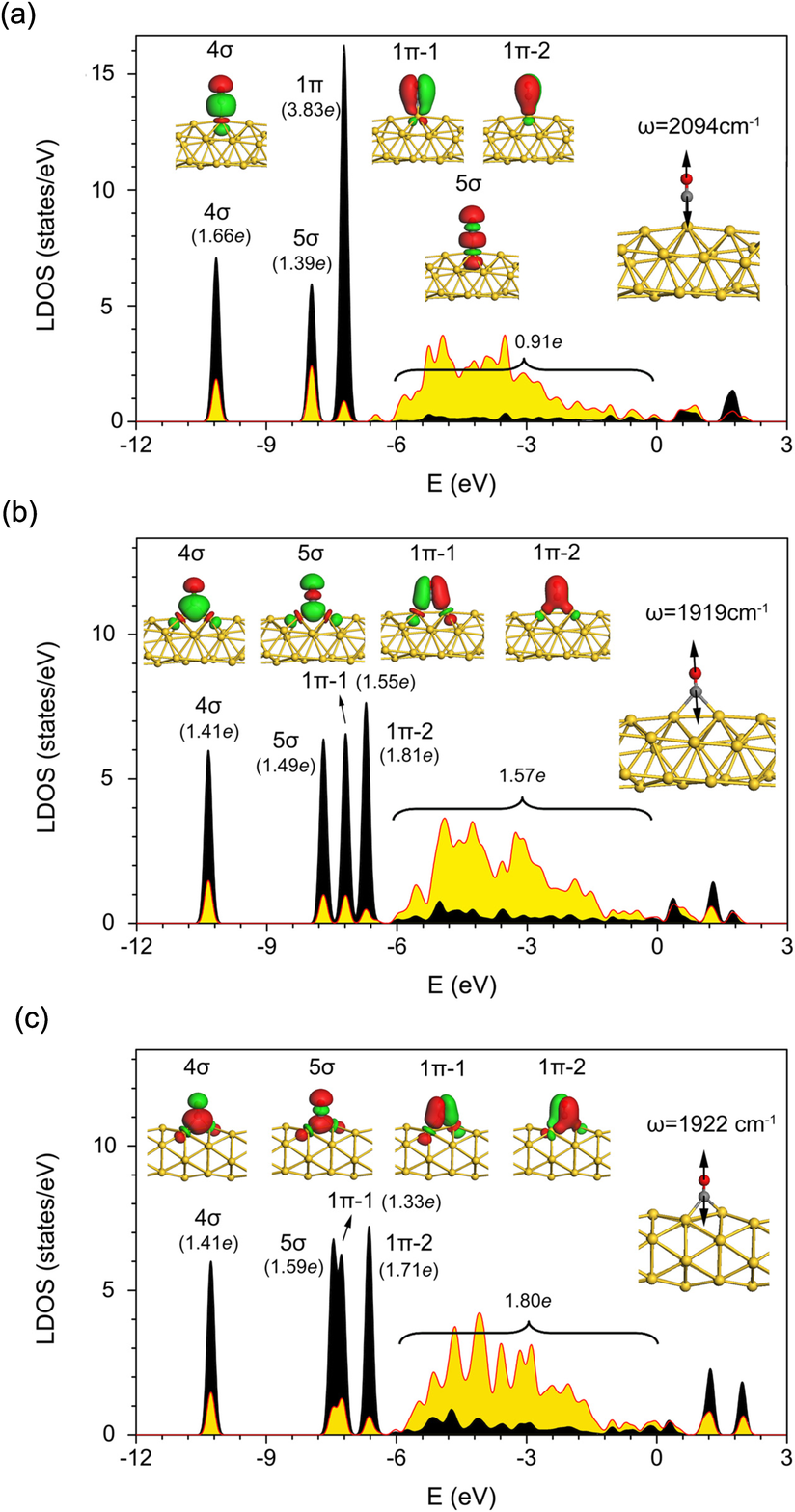}\\
  \caption{(Color online) LDOS and orbitals of CO adsorbed on Au(5,3) at top(a), B1(b) sites and Au(5,5) at B1(c) site. The LDOS projected on the CO molecule is shown in black and that projected on the gold atom with the shortest distance from the C atom is colored yellow. The populations of  CO orbitals are calculated through integrating over the relevant peaks. Included also are the vibrational frequencies of stretching mode for different adsorption configurations. The softening frequencies are evidence of the back-donation process during adsorption. Notice that the frequency of free CO molecule is calculated as 2128 $cm^{-1}$}\label{Fig.3}
\end{figure}

\begin{figure}
  \includegraphics[width=9.0 cm]{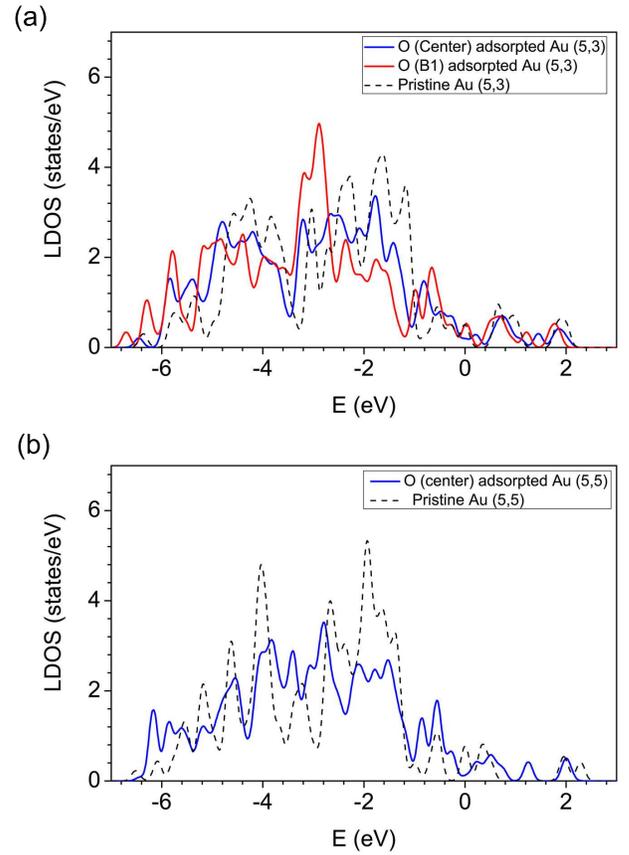}\\
  \caption{(Color online) Comparison of Au states after O atomic adhesion with those of the pristine (5,3)(a) and (5,5)(b) Au tubes. For the adsorbed tubes, LDOS is projected onto Au atom with the shortest distance from the O atom.}\label{Fig.4}
\end{figure}

  The significantly different backdonation strength between the top site and B1 site adsorption implies a way to distinguish these two almost degenerate adsorption configurations of CO on Au(5,3) in experiment. The more the CO anti-binding $2\pi^{*}$ orbital is backdonated, the weaker the CO bond will be. The strength of the CO bond can be seen from the bond length and also the vibrational frequency of the stretching mode of the molecule, the latter of which can be measured in experiment. Indeed, the CO bond length of the B1 adsorption, 1.165 ${\AA}$, is larger than that of the top adsorption, 1.151 ${\AA}$ due to the stronger backdonation. Note that the bond length of free CO molecule is 1.128 ${\AA}$ (Our calculated value of 1.146 ${\AA}$). The vibrational frequency of the CO stretching mode is calculated to be 2094 $cm^{-1}$ for top adsorption and 1919 $cm^{-1}$ for B1 adsorption. The difference between these two vibrational frequencies should be able to be seen in experiment. As a test of the method of calculating vibrational spectra, for the same mode of the free CO molecule, our calculations give the frequency of 2128 $cm^{-1}$, which is in good agreement with the experimental value of 2140 $cm^{-1}$.\cite{H.-J. Freund}

  The backdonation of $2\pi^{*}$ orbital, and the splitting of $1\pi-1$ and $1\pi-2$, are also observed for the lowest energy adsorption of CO on Au(5,5), the B1 adsorption (Fig. 3c). Here, the nearly 1 eV of the splitting of two $1\pi$ orbitals is much larger than that of the B1 adsorption of Au(5,3) (about 0.5 eV). The CO bond length in this case is calculated to be 1.172 ${\AA}$, and the corresponding vibrational frequency of the CO stretching mode is 1922 $cm^{-1}$, about the same as that of the B1 adsorption on Au (5,3).

  In Fig. 4, we show the LDOS of the Au atom binded with the O before and after O adsorption. Again, for Au(5,3), there are two nearly degenerate adsorption configurations, the center and the B1 adsorption, and for Au(5,5), the most stable configuration occurs when the O atom binds to the center position. For all cases, the O adsorption causes great changes of LDOS of the binded Au atom, indicating the strong interaction between the O and the Au tube.

  \subsection{Conductance of CO/O absorbed Au tubes}

\begin{figure}
  \includegraphics[width=9.0 cm]{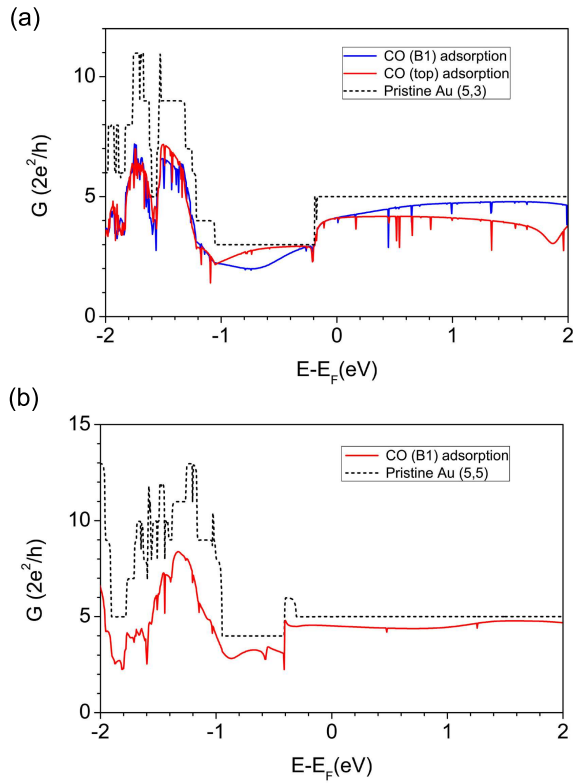}\\
  \caption{(Color online) Quantum conductance spectrum for CO adsorbed Au(5,3)(a) and Au(5,5)(b). The conductance spectra of pristine tubes(dashed lines) are also given for comparison}\label{Fig.5}
\end{figure}

  Now, we turn to the analysis of conductance of chemically modified Au tubes. It has been reported that for an Au monoatom chain that has one unit of conductance, the adsorption of a single CO molecule can make a sharp modification of conductance of Au monochain from 1 $G_{0}$ to 0 after adsorption at top site.\cite{A. Calzolari} It would be interesting to see how the adsorption of CO or O modify the conductance of Au (5,3) or (5,5) tubes that essentially are the aggregation of five strands of Au atoms.

  The calculated conductance spectra of the pristine and CO adsorbed Au (5,3) and Au (5,5) are shown in Fig. 5. The conductance of both pristine tubes at Fermi energy is 5 $G_{0}$, in agreement with previous studies. \cite{R. T. Senger} For Au(5,3), the CO adsorption decreases the conductance at Fermi energy of the tube by 0.9 $G_{0}$ for both top and B1 adsorptions, suggesting that Au(5,3) may be used as a chemical sensor for CO molecules. For Au(5,5), the lowest-energy configuration of CO adsorption decreases the conductance at Fermi energy of the tube by 0.5 $G_{0}$. Compared to CO, effects of O atom on transport properties are much more pronounced due to the stronger interaction between the O atom and the Au tube (Fig. 6). For both tubes, the O adsorption causes the drop of the conductance at Fermi energy by around 2 $G_{0}$.

\begin{figure}
  \includegraphics[width=9.0cm]{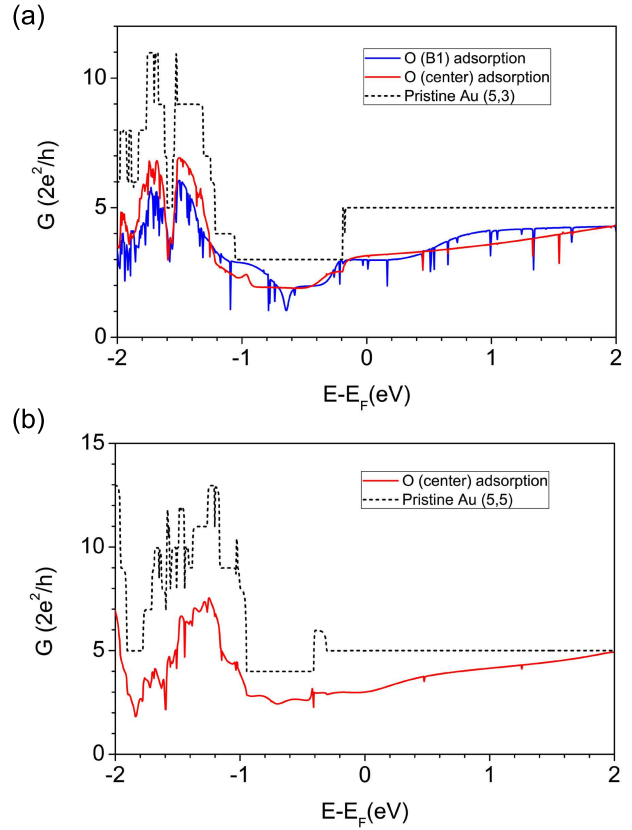}\\
  \caption{(Color online) Quantum conductance spectrum for oxygen adsorbed Au(5,3)(a) and Au(5,5)(b). The conductance spectra of pristine tubes(dashed lines) are also given for comparison}\label{Fig.5}
\end{figure}

  \subsection{Defects effect on conductance of Au tubes}

  Finally, we discuss the effects of two types of defects, the Au monovacancy and the defect arising from the adhesion of one extra Au atom, on conductance of Au tubes. In Table I, we give the adsorption energies of a single Au atom on Au(5,3) (2.28eV@center and 2.27 eV@B2) and Au(5,5) (2.69eV@center). The formation energies of a monovacancy are calculated to be 3.32 and 3.52 eV for Au(5,3) and Au(5,5) respectively, which are rather small compared to those of CNTs.\cite{B. Biel} The high adsorption energies of one Au atom and relatively low formation energies of monovancy suggest that these two types of defects are likely to occur in the fabrication of Au tubes.

\begin{figure}
  \includegraphics[width=9.0 cm]{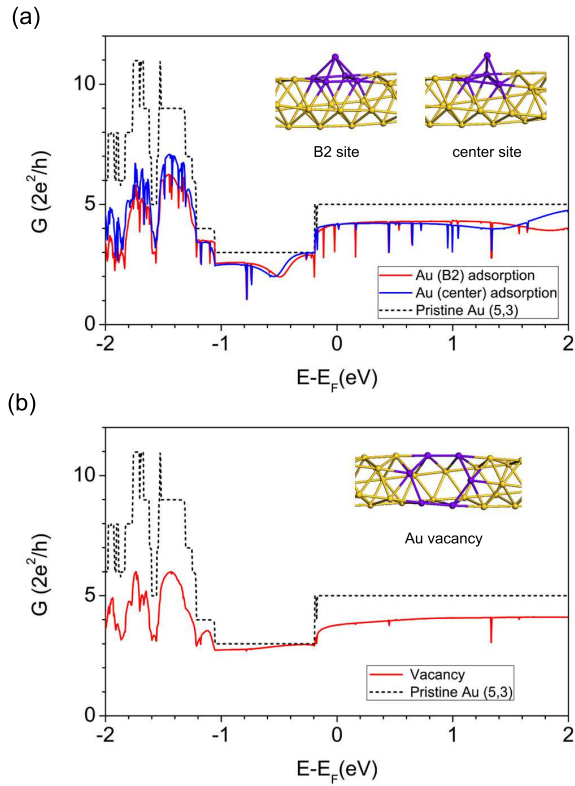}\\
  \caption{(Color online) Conductance as a function of energy for defective Au(5,3) tube with defects arising from Au adhesion(a) and monovacancy(b) on the tube. The relaxed defective structures are given as the insets, where the atoms around the defective site are highlighted  by violet balls. It can be seen that the Au adatom distorts the tube differently at different adsorption sites.}\label{Fig.6}
\end{figure}

\begin{figure}
  \includegraphics[width=9.0cm]{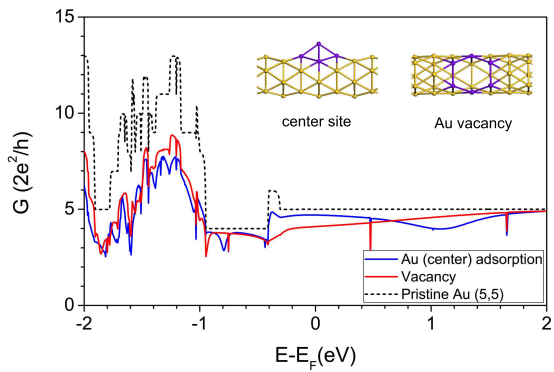}\\
  \caption{(Color online) Conductance as a function of energy for defective Au(5,5) tube with defects arising from Au adhesion(blue line) and monovacancy(red line) on the tube. The relaxed defective structures are given as the insets, where the atoms around the defective site are highlighted by violet balls.}\label{Fig.7}
\end{figure}

  The conductance spectra and optimized structures of defective Au(5,3) are given in Fig. 7. For cases of the extra Au atom adsorbed on the tube (B2 and center sites), after relaxation, pyramid-like structures are formed as shown in Fig. 7a. The conductance spectra of both two adsorption configurations show very similar behaviors, and at Fermi energy, the conductance decreases by about 0.8 $G_{0}$ for both configurations. For Au monovacancy, the defect decreases the conductance of the tube at Fermi energy about 1.2 $G_{0}$ (Fig. 7b). Similarly, the Au monovacancy on Au(5,5) causes a 0.9 $G_{0}$ of drop of conductance at Fermi energy (Fig. 8). The Au adatom on Au(5,5) only leads to a decrease of 0.3 $G_{0}$ of the tube as we can see from Fig. 8.

\section{CONCLUSIONS}

  We have investigated via the first principles method based on DFT the influence of adsorbates (CO molecule and O atom) and defects on the the electronic and transport properties of Au(5,3) and Au(5,5) nanotubes. For Au(5,3), we found that there are two nearly degenerate stable adsorption configurations(top site and B1 site adsorption) for both CO molecule and O atom. For the case of CO, for both tubes, HOMO and LUMO states of CO contribute the most to the chemisorption. For one of these two degenerate configurations of Au(5,3), the B1 site adsorption, and the stable adsorption configuration of Au(5,5), the $1\pi$ orbital is also involved in the bonding between CO and Au tube. The charge transfer between the CO molecule and both Au tubes is dominated by so-called backdonation process in which the electrons transferred from Au $d$ bands to the CO anti-binding $2\pi^{*}$ orbital. We also predicted a significant difference between the vibrational frequencies of the CO molecule for two degenerate stable adsorption configurations of Au(5,3), which may be used to distinguish these two configurations in experiments.

  After CO adsorption, the conductance of Au(5,3) decreases by 0.9 $G{_0}$, and the conductance of Au(5,5) decreases by approximately 0.5 $G{_0}$. For O adsorbed Au tubes, O atoms strongly interact with Au tubes, leading to around 2 $G{_0}$ of drop of conductance for both Au tubes. These results may have implications for Au-tube based chemical sensing. When a monovacancy defect is present, we found that for both tubes, the conductance decreases by around 1 $G{_0}$. Another type of defect arising from the adhesion of one Au atom is also considered. For this case, it is found that the defect decreases the conductance by nearly 1 $G{_0}$ for Au(5,3) tube, and the decrease of the conductance is around 0.3 $G{_0}$ for Au(5,5) tube.

Acknowledgement: This work was supported by NUS Academic Research Fund (Grant Nos:
R-144-000-237-133 and R-144-000-255-112). Computations were performed
at the Centre for Computational Science and Engineering at NUS.


\begin{thebibliography}{0}
\expandafter\ifx\csname natexlab\endcsname\relax\def\natexlab#1{#1}\fi
\expandafter\ifx\csname bibnamefont\endcsname\relax
  \def\bibnamefont#1{#1}\fi
\expandafter\ifx\csname bibfnamefont\endcsname\relax
  \def\bibfnamefont#1{#1}\fi
\expandafter\ifx\csname citenamefont\endcsname\relax
  \def\citenamefont#1{#1}\fi
\expandafter\ifx\csname url\endcsname\relax
  \def\url#1{\texttt{#1}}\fi
\expandafter\ifx\csname urlprefix\endcsname\relax\def\urlprefix{URL }\fi
\providecommand{\bibinfo}[2]{#2}
\providecommand{\eprint}[2][]{\url{#2}}

\end{thebibliography}


\begin{thebibliography}{}
\bibitem{Y. Kondo}
Y. Kondo and K. Takayanagi, Science 289, 606 (2000).
\bibitem{STM}
H. Ohnishi, Y. Kondo, and K. Takayanagi, Nature 395, 780 (1998).
\bibitem{Break Junction}
N. Agra\"{\i}t, A. L. Yeyati, and J. M. van Ruitenbeek, Phys. Rep. 377, 81 (2003).
\bibitem{Y. Oshima_prl}
Y. Oshima, A. Onga, and K. Takayanagi, Phys. Rev. Lett. 91, 205503 (2003).
\bibitem{N. Nilius_science}
N. Nilius, T.M. Wallis, and W. Ho, Science 297, 1853 (2002).
\bibitem{N. Nilius_prl}
N. Nilius, T.M. Wallis, and W. Ho, Phys. Rev. Lett. 90, 186102 (2003).
\bibitem{A. I. Yanson}
A. I. Yanson, G. Rubio Bollinger, H. E. van den Brom, N. Agra\"{\i}t, and J. M. van Ruitenbeek, Nature 395, 783 (1998)
\bibitem{R. T. Senger}
R. T. Senger, S. Dag, and S. Ciraci, Phys. Rev. Lett. 93, 196807 (2004).

\bibitem{Spiral_Current}
T. Ono, and K. Hirose, Phys. Rev. Lett. 94, 206806 (2005).
\bibitem{D. Z. Manrique}
D. Z. Manrique, J. Cserti, and C. J. Lambert, Phys. Rev. B 81, 073103 (2010).
\bibitem{M. A. Sanchez-Castillo}
M. A. Sanchez-Castillo, C. Couto, W. B. Kim, and J. A. Dumesic, Angew. Chem. Int. Ed. 43, 1140 (2004).
\bibitem{W. An}
W. An, Y. Pei, and X. C. Zeng, Nano Lett. 8, 195 (2008).
\bibitem{Zhang_Nanowire}
C. Zhang, R. N. Barnett, U. Landman, Phys. Rev. Lett.  100, 046801 (2008).
\bibitem{S. Wippermann}
S. Wippermann, N. Koch, and W. G. Schmidt, Phys. Rev. Lett. 100, 106802 (2008)
\bibitem{A. Calzolari}
A. Calzolari, C. Cavazzoni, and M. B. Nardelli, Phys. Rev. Lett. 93, 096404 (2004).
\bibitem{ZMpaper}
M. Zhou, A. Zhang, Z. Dai, C. Zhang, and Y. P. Feng, J. Chem. Phys. 132, 194704 (2010)


\bibitem{VASP_jcm}
G. Kresse, Furthm¨¹ller, J. Comput. Mater. Sci. 6, 15 (1996).
\bibitem{VASP_prb}
G. Kresse, Furthm¨¹ller, Phys. Rev. B 54, 11169 (1996).
\bibitem{ATK_1}
M. Brandbyge, J. -L. Mozos, P. Ordej¨®n, J. Taylor, K. Stokbro, Phys. Rev. B 65, 165401 (2002).
\bibitem{ATK_2}
J. M. Soler, E. Artacho, J. D. Gale, A. Garc¨ªa, J. Junquera, P. Ordej¨®n, D. S¨¢nchez-Portal, J. Phys. Condens. Matter 14, 2745 (2002).
\bibitem{ATK_3}
J. Taylor, H. Guo, J. Wang, Phys. Rev. B 63, 245407 (2001).
\bibitem{PBE_GGA}
J. Perdew, K. Burke, M. Ernzerhof, Phys. Rev. Lett. 77, 3865 (1996).
\bibitem{DMol_1}
B. Delley, J. Chem. Phys. 92, 508 (1990).
\bibitem{DMol_2}
B. Delley, J. Chem. Phys. 113, 7756 (2000).

\bibitem{Landman_science}
B. Yoon, H. H\"{a}kkinen, U. Landman, A. S. W\"{o}rz, J.-M. Antonietti, S. Abbet, K. Judai, U. Heiz, science 307, 403 (2005).
\bibitem{G. Blyholder}
G. Blyholder, J. Phys. Chem. 68, 2772 (1964).

\bibitem{M. Kiguchi}
M. Kiguchi, D. Djukic, and J. M. van Ruitenbeek, Nanotechnology 18, 035205 (2007).
\bibitem{H.-J. Freund}
H.-J. Freund and G. Pacchioni, Chem. Soc. Rev. 37, 2224(2008).
\bibitem{B. Biel}
B. Biel, F. J. Garc\'{\i}a-Vidal, \'{A}. Rubio, and F. Flores, J. Phys. Condens. Matter, 20, 294214 (2008).

\end{thebibliography}
\end{document}